\documentclass[prd,aps,showpacs,nofootinbib,nobibnotes,twocolumn]{revtex4}
\usepackage{psfig}

\newcommand{\ETAL}{{\it et al.}}

\newcommand{\MAT}{{\rm m}}
\newcommand{\LSS}{{\scriptscriptstyle \rm LSS}}

\newcommand{\ALM}[2]{a_{{#1}\,{#2}}}
\newcommand{\CLMLPMP}[4]{C_{{#1}\,{#2}}^{{#3}\,{#4}}}

\newcommand{\UUNIT}[2]{\,{\mbox{#1}^{#2}}}

\newcommand{\YXKLM}[4]{{\cal Y}^{[{#1}]}_{{#2}\,{#3}\,{#4}}}
\newcommand{\UPSTKS}[3]{\Upsilon^{[{#1}]}_{{#2}\,{#3}}}
\newcommand{\XITKSLM}[5]{\xi^{[{#1}]\,{#3}}_{{#2}\,{#4}\,{#5}}}

\newcommand{\KRON}[2]{\delta_{{#1}\,{#2}}}

\newcommand{\bb}{\bf}
\newcommand{\STR}{{\bb S}^3}
\newcommand{\CHN}{\bar \chi}

\begin{document}
\title{Cosmic microwave background constraints on lens spaces}

\author{Jean-Philippe Uzan}
\email{uzan@iap.fr}
\affiliation{Institut d'Astrophysique de Paris, GR$\varepsilon$CO, FRE
             2435-CNRS, 98bis boulevard Arago, 75014 Paris, France \\
             Laboratoire de Physique Th\'eorique, CNRS-UMR 8627,
             Universit\'e Paris Sud, B\^atiment 210, F--91405 Orsay
             c\'edex, France}

\author{Alain Riazuelo}
\email{riazuelo@iap.fr}
\affiliation{Service de Physique Th{\'e}orique,
             CEA/DSM/SPhT, Unit{\'e} de recherche associ{\'e}e au
             CNRS, CEA/Saclay F--91191 Gif-sur-Yvette c{\'e}dex,
             France}

\author{Roland Lehoucq}
\email{lehoucq@cea.fr} \affiliation{CEA-Saclay,
  DSM/DAPNIA/Service d'Astrophysique,
  F--91191 Gif-sur-Yvette c\'edex, France}

\author{Jeffrey Weeks}
\email{weeks@northnet.org}
\affiliation{15 Farmer St.,  Canton  NY  13617-1120, USA}

\date{22 October 2003}
\begin{abstract}
\noindent This article describes the Cosmic Microwave Background
anisotropies expected in a closed universe with the topology of a
lens space $L(p,q)$ and with density parameter $\Omega_0$ close
to~$1$.  It provides the first simulated maps for such spaces
along with their corresponding power spectra. In spite of our
initial expectations that increasing $p$ (and thus decreasing the
size of the fundamental domain) should suppress the quadrupole, we
found just the opposite:  increasing $p$ elevates the relative
power of the low multipoles, for reasons that have since become
clear.  For $\Omega_0 = 1.02$, an informal ``by eye'' examination
of the simulated power spectra suggests that $p$ must be less than
15 for consistency with WMAP's data, while geometric
considerations imply that matching circles will exist (potentially
revealing the multi-connected topology) only if $p > 7$. These
bounds become less stringent for values of $\Omega_0$ closer to 1.
\end{abstract}
\maketitle

\section{Introduction} 

High resolution full sky cosmic microwave background (CMB) maps are
now available from the Wilkinson Microwave Anisotropy Probe (WMAP)
experiment~\cite{wmap}, offering an unprecedented opportunity to probe
the global structure of our universe (see Refs.~\cite{revues} for
reviews). This article discusses expected CMB anisotropies in
spherical spaces, providing simulated CMB maps and their corresponding
power spectra produced using the methods of our companion
papers~\cite{rulw,uwrl}.

The recent WMAP results interestingly indicate that a closed
universe seems to be marginally preferred~\cite{wmap}. In
particular, with a prior on the Hubble constant, one gets a
density parameter $\Omega_0 = 1.03 \pm 0.05$, while further
including type Ia supernovae data leads to $\Omega_0 = 1.02 \pm
0.02$.  Moreover, the WMAP angular correlation function seems to
lack signal on scales larger than 60 degrees~\cite{wmap}. This may
indicate a possible discreteness and a cutoff in the initial power
spectrum, as would be expected from a multiconnected topology.
Contrary to the standard lore, we will see that for lens spaces
$L(p,q)$ the topology leads to an increase of the CMB power on
large angular scales. This boost of power is related to the fact
that space becomes more and more anisotropic as $p$ increases. The
same effect also appears for (flat) slab and chimney
spaces~\cite{uwrl,is} but not in a well-proportioned space as the
Poincar\'e dodecahedral space~\cite{PDS}.  In addition,
cosmological data indicate that $|\Omega_0 - 1|$ is small, which
has important implications concerning the observability of the
topological structure of our universe~\cite{wlu02}.

It is unlikely that we could detect a compact hyperbolic space if
$\Omega_0 - 1$ is small and negative, because, for a generic observer,
the topology scale is comparable to the curvature radius or
longer. Only an observer lucky enough to live near a short closed
geodesic could detect topology~\cite{weeks}.

In flat universes, the topological scale is completely independent of
the horizon radius, because Euclidean geometry has no preferred scale
and admits similarities. There is no fundamental reason for the
topology scale to be less than the horizon radius but still large
enough to accommodate the lack of obvious local periodicity. Compact
flat universes have nevertheless been studied extensively. In
particular, it was shown on the basis of the COBE data that, for a
vanishing cosmological constant, no more than $8$ copies of the
fundamental cell within our horizon are allowed~\cite{tore}. A
non-vanishing cosmological constant relaxes the constraint to $49$
copies~\cite{inoue01} if $\Omega_\Lambda = 0.9$ and $\Omega_\MAT =
0.1$. A catalog of maps for all flat spaces (compact, chimney and
slab) will soon be available~\cite{uwrl}.

In spherical spaces the topology scale is tied to the curvature
radius, but as the topology gets more complicated, the typical
shortest distance between two topological images decreases. So, no
matter how flat the universe, all but a finite number of spherical
topologies remain detectable.

\begin{figure}
\centerline{\psfig{file=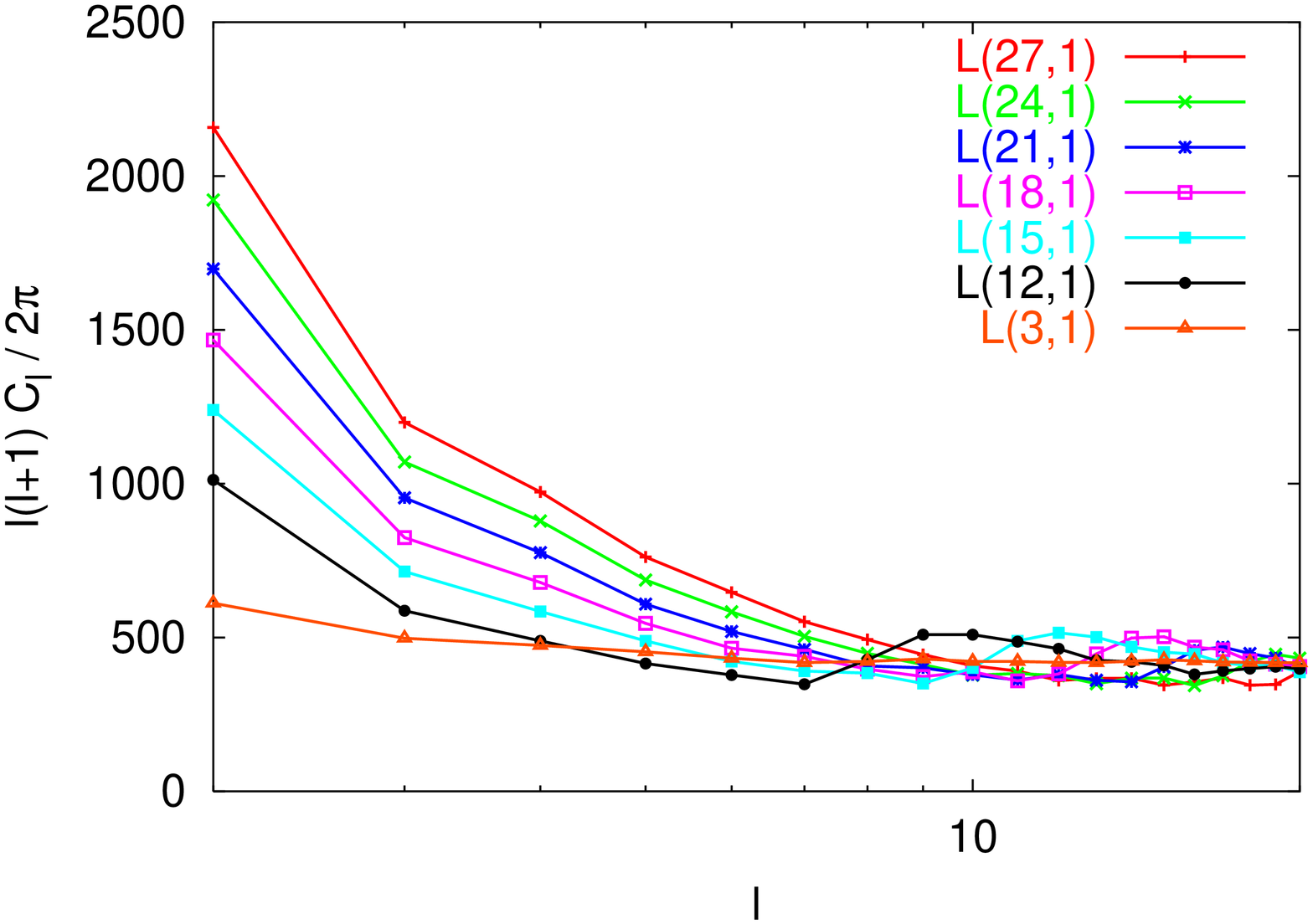,width=3.0in,angle=270}}
\centerline{\psfig{file=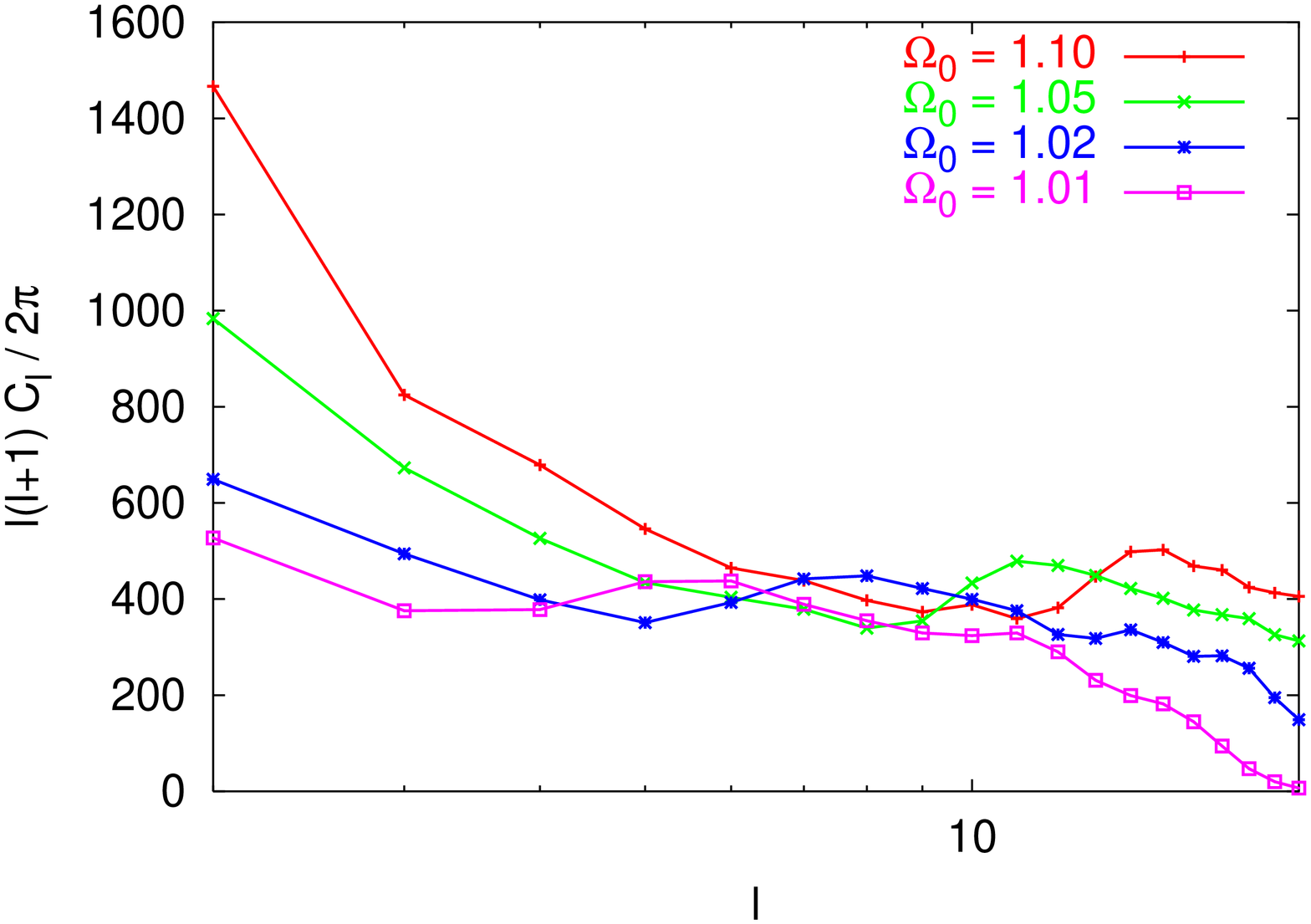,width=3.0in,angle=270}}
\caption{The angular power spectrum for the lens spaces $L (p, 1)$
with $p = 3, 12, 15, 18, 21, 24, 27$ for $\Omega_0 = 1.1$.  The
relative power of the large angular scale increases with $p$
[top]. Bottom panel shows lens space $L (18,1)$ and $\Omega_0 =
1.01$, $1.02$, $1.05$, $1.1$. Here, the increase of power on large
scales is due to the usual Integrated Sachs-Wolfe effect. (The
cutoff at smaller angular scales is unphysical and is simply due
to the fact that we used an insufficient number of small
wavelength modes.)} \label{fig1}
\end{figure}

\section{Spherical multiconnected spaces} 

Three-dimensional spherical spaces were originally classified by
Threlfall and Seifert~\cite{ThrelfallSeifert} in 1930. The
classification was recently revisited in terms of single, double and
linked action manifolds by Gausmann \ETAL~\cite{glluw}.  Borrowing
from Thurston's approach~\cite{thurston}, we used the fact that any
finite group of unit quaternions determines a fixed point free group
$\Gamma$ of isometries of $\STR$, which then serves as the holonomy
group of a multiply connected spherical space. The spaces arising in
this way are called {\it single action} spaces and are in one-to-one
correspondence with the finite subgroups of $\STR$, thought of as the
group of all unit length quaternions.  The finite subgroups of $\STR$
are the cyclic groups $Z_n$, the binary dihedral groups $D^*_m$, the
binary tetrahedral, octahedral and icosahedral groups, respectively of
order $n$, $4m$, 24, 48 and 120. In a {\it double action} space, two
groups of relatively prime order act simultaneously so that $\Gamma$
is the product of a cyclic group by either a cyclic or or a binary
polyhedral group. {\it Linked action} spaces are similar to double
action spaces except that the orders of the factors are not relatively
prime and only certain elements of one factor are allowed to act
simultaneously with a given element of the other factor. In all cases
the volume of the space $\STR / \Gamma$ is the volume of the 3-sphere
$\STR$ divided by the order $|\Gamma|$ of the holonomy group.

{\it Lens spaces} can be single action, double action, or linked
action.  A lens space's fundamental domain is constructed by
identifying the two faces of a lens shaped solid with a $2 \pi q /
p$ rotation, for relatively prime integers $p$ and $q$ such that
$0 < q < p$.  The result is the lens space $L (p, q)$. Exactly $p$
copies of the fundamental domain tile the 3-sphere, their faces
lying on great 2-spheres filling a hemisphere of each, just as the
2-dimensional surface of an orange may be tiled with $p$ sections
of orange peel, meeting along meridians spaced $2 \pi / p$ apart.
Different lens spaces may share the same abstract holonomy group;
for example $L (5, 1)$ and $L (5, 2)$ both have $\Gamma \approx
Z_5$ even though the group acts differently in each case.
Non-cyclic fundamental groups, by contrast, act in a unique way.

\begin{figure}
\centerline{\psfig{file=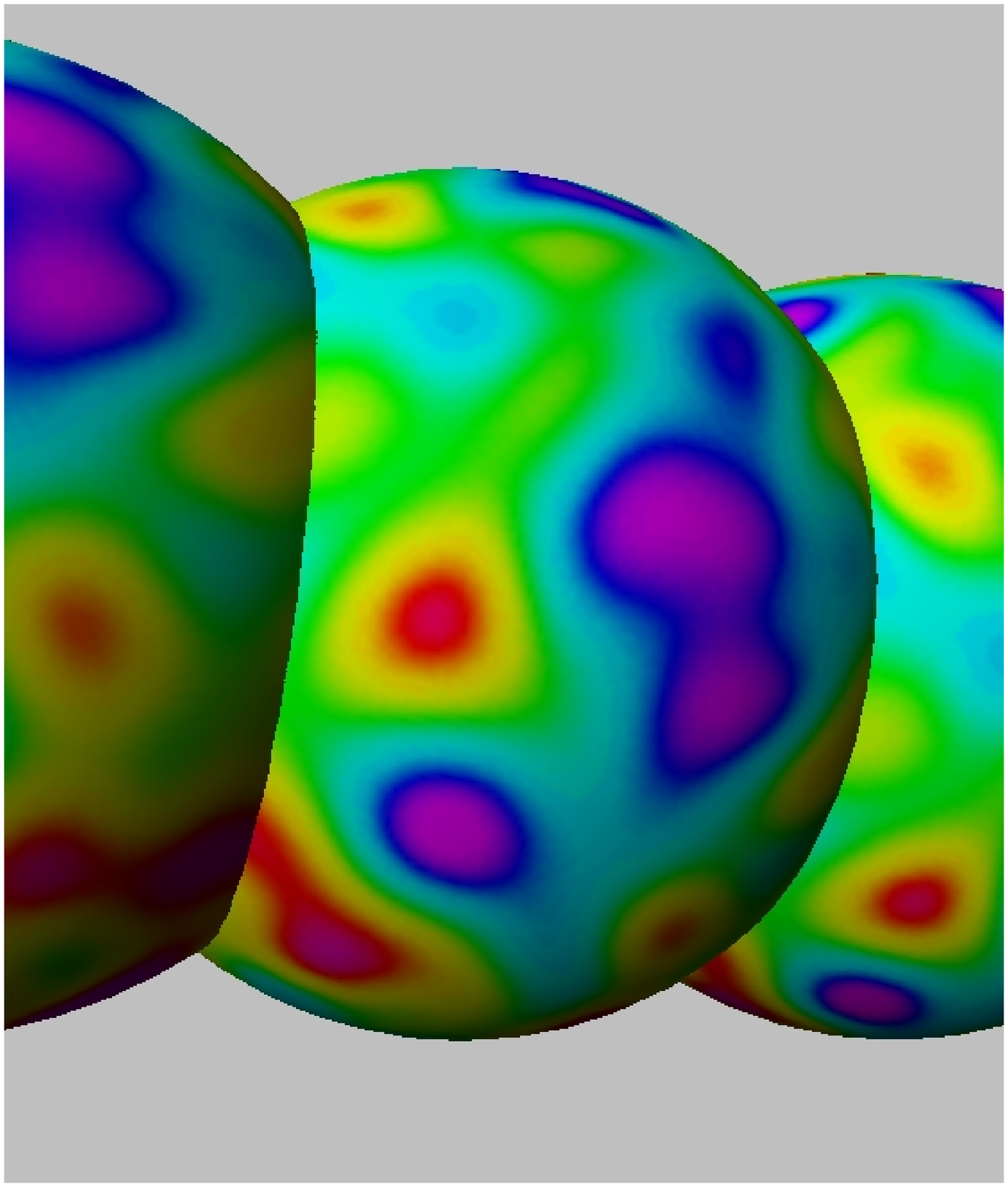,width=1.75in}
            \psfig{file=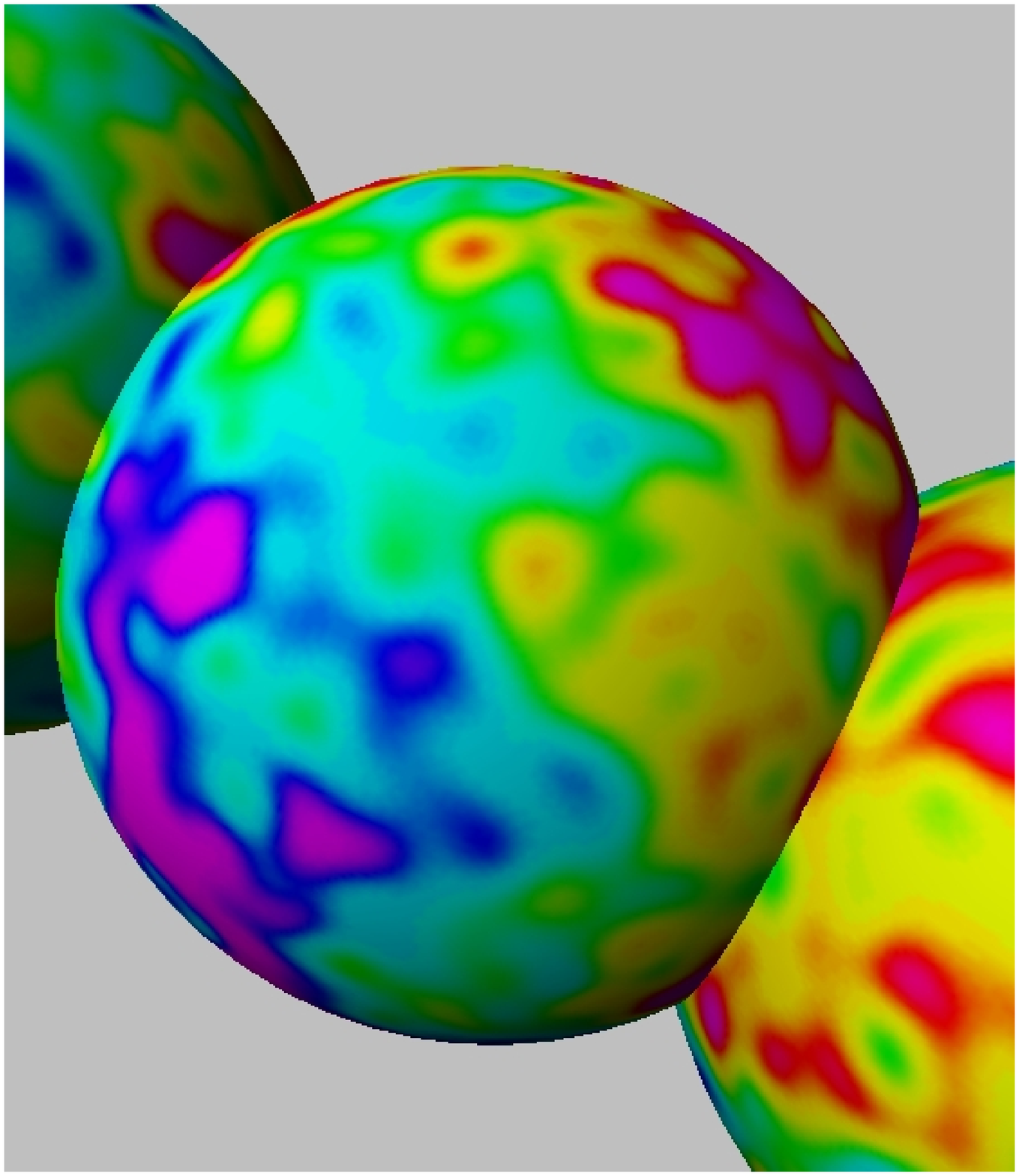,width=1.75in}}
\caption{Simulated maps for the lens space $L (12, 1)$.  Matching
circles are opposite but match with a twist of $2 \pi / p$.  When
$\Omega_0 = 1.02$ [left] there is a single pair of circles; when
$\Omega_0 = 1.1$ [right] there are three pairs; only the smallest
pair of circles is shown.}
\label{fig2}
\end{figure}
\begin{figure}
\centerline{\psfig{file=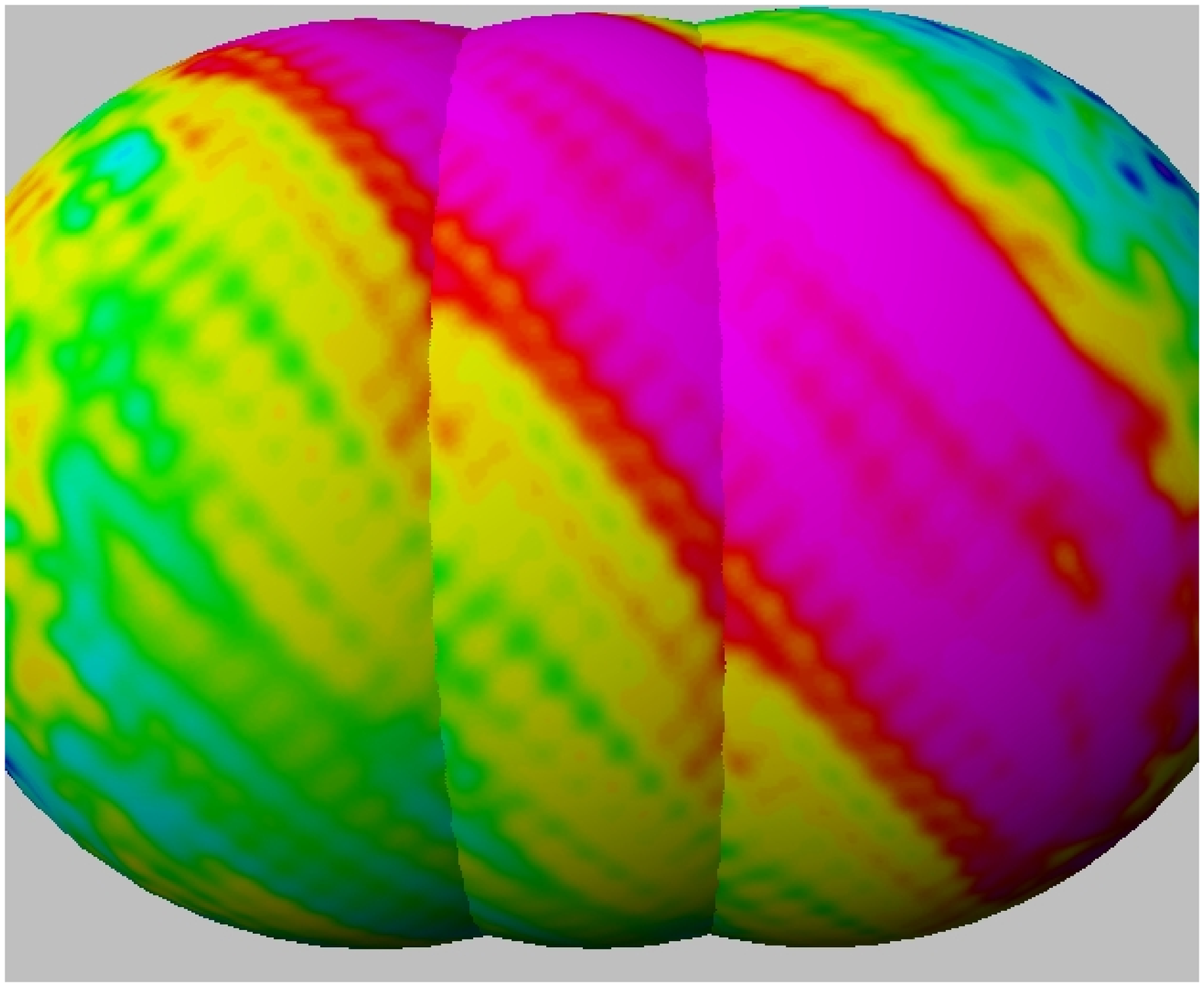,angle=270,width=1.75in}
            \psfig{file=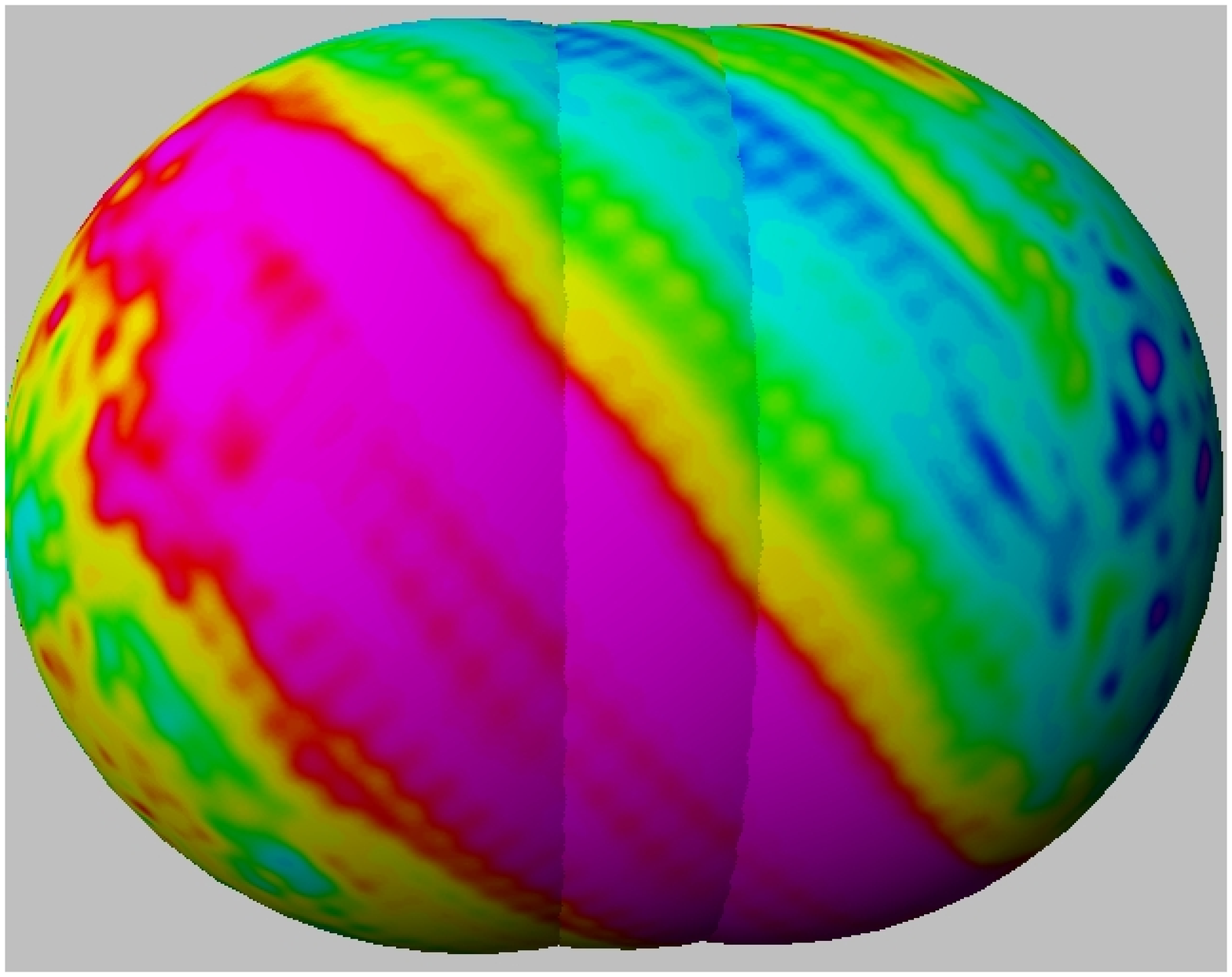,angle=270,width=1.75in}}
\caption{Same as Fig.~\ref{fig2} but in the extreme case of $L
(100,1)$ for $\Omega_0 = 1.1$.  The short period ($2 \pi / 100$) makes
the symmetry along Clifford parallels obvious to the eye.}
\label{fig3}
\end{figure}

Topology breaks the global isotropy and possibly the global
homogeneity of the universe, the only exception being projective
space (see Ref.~\cite{lwugl,souradeep}). Consequently, the CMB
temperature angular correlation function will depend on the two
directions of observation, not only on their relative angle, and
possibly on the position of the observer as well. This induces
correlations between the $\ALM{\ell}{m}$ of different $\ell$ and
$m$. Such correlations are hidden when one considers only the
angular correlation function and its coefficients, $C_\ell$, in a
Legendre polynomial decomposition, because they pick up only the
isotropic part and are therefore a poor indicator of the topology.
The correlation matrix $\CLMLPMP{\ell}{m}{\ell'}{m'} \equiv \left<
\ALM{\ell}{m} \ALM{\ell'}{m'}^* \right>$ encodes all the
topological properties of the CMB; see Ref.~\cite{rulw} for
details. Conversely, in a simply-connected space it reduces to the
usual formula $\CLMLPMP{\ell}{m}{\ell'}{m'}= C_\ell \KRON{\ell}{\ell'}
\KRON{m}{m'}$.

\begin{figure}
\centerline{\psfig{file=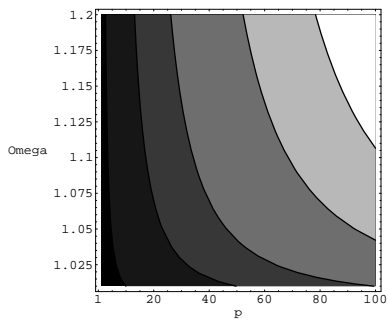,width=2.0in}}
\caption{Number $N$ of matched circle pairs for a lens space $L
(p, 1)$, for $p$ between $1$ and $100$ and $\Omega_0$ between
$1.01$ and $1.2$. The contour lines are, from left to right, for
$N = 1$, $5$, $10$, $20$, $30$.} \label{fig1c}
\end{figure}

\begin{figure}
\centerline{\psfig{file=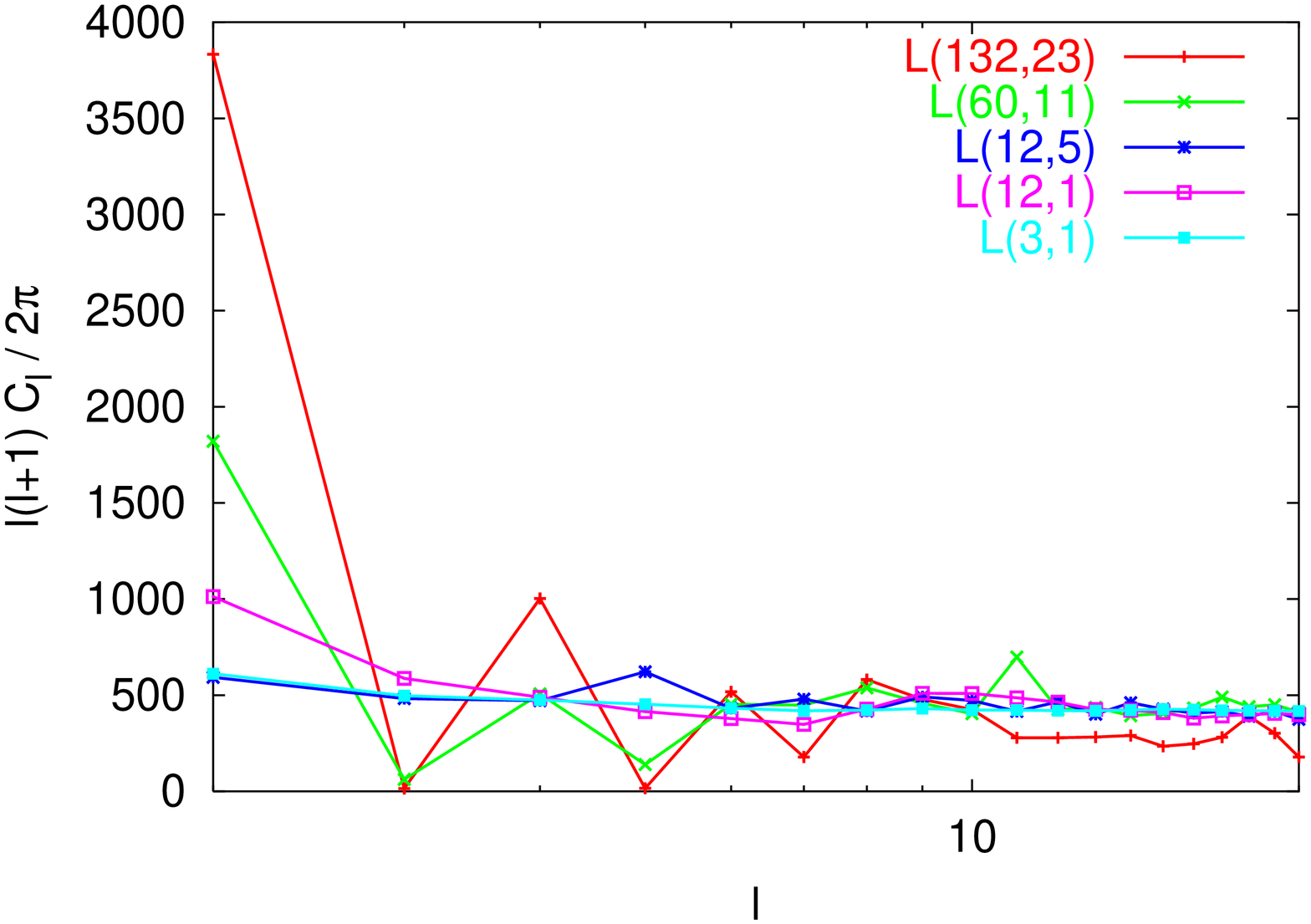,width=3.0in,angle=270}}
\centerline{\psfig{file=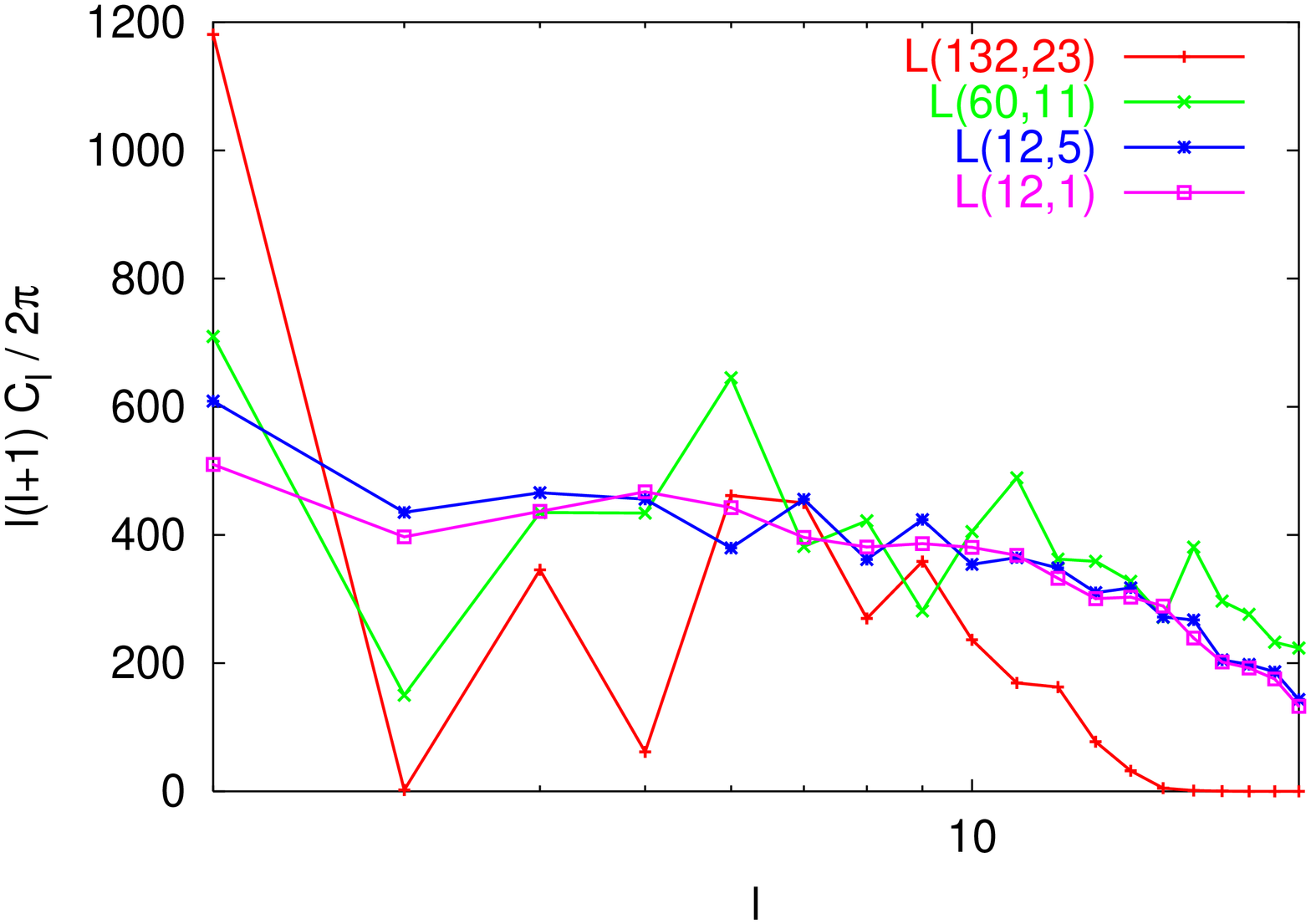,width=3.0in,angle=270}}
\caption{Influence of the cyclic factors: $L (12, 5)$ and $L (12,
1)$ have the same volume but the former typically does not give
rise to matching circles; $L (60, 11)$ and $L (132, 23)$ both have
a cyclic factor $Z_{12}$ just like $L (12, 1)$ but have smaller
volumes because of an additional factor of $Z_5$ and $Z_{11}$,
respectively. Data are shown for $\Omega_0 = 1.1$ [top] and
$\Omega_0 = 1.02$ [bottom].} \label{fig1d}
\end{figure}

In a companion paper~\cite{rulw}, we showed that CMB maps with a
topological signal can be confidently simulated for flat and
spherical spaces. Topology does not affect local physics and
enters only via the boundary conditions of the functions defined
on the multi-connected space.  Technically, the equations of the
cosmological perturbations reduce to a set of coupled differential
equations involving the Laplacian operator. They are conveniently
solved in Fourier space but this requires the determination of the
eigenmodes $\UPSTKS{\Gamma}{k}{}$ with the correct boundary
conditions. In $\STR$, the eigenmodes $\YXKLM{\STR}{k}{\ell}{m}$
are labelled by an integer $\nu$, related to $k$ by $k = (\nu + 1)
\sqrt{K}$ where $K$ is the spatial curvature, and each eigenvalue
$\nu (\nu + 2) K$ has a multiplicity $(\nu + 1)^2$. The eigenmodes
$\UPSTKS{\Gamma}{k}{}$ can be expressed in terms of the eigenmodes
of $\STR$ as $\UPSTKS{\Gamma}{k}{s}= \sum_{\ell, m}
\XITKSLM{\Gamma}{k}{s}{\ell}{m} \YXKLM{\STR}{k}{\ell}{m}$, so that
all the topological information is encoded in the coefficients
$\XITKSLM{\Gamma}{k}{s}{\ell}{m}$, where $s$ indexes the
eigenmodes sharing a common value of $k$. One finds that the Weyl
formula holds for spherical spaces, so a space of order $|\Gamma|$
has approximately $N^{\scriptscriptstyle{[\Gamma]}} \sim
N^{\scriptscriptstyle{[\STR]}} / |\Gamma|$ modes~\cite{glluw}. It
follows that $|\Gamma| C_\ell^{\scriptscriptstyle{[\Gamma]}} \to
C_\ell^{\scriptscriptstyle{[\STR]}}$ when $\ell$ becomes
large~\cite{lwugl} (see also Ref.~\cite{rulw} for a numerical
check of this behaviour for a torus).

\section{Maps, power spectra and constraints} 

For lens and prism spaces the coefficients
$\XITKSLM{\Gamma}{k}{s}{\ell}{m}$ were recently computed analytically
using a toroidal coordinate system~\cite{luw02}. CMB codes generally
use spherical coordinates but the change of coordinates has been
handled~\cite{rulw,lwugl}.

For a homogeneous lens space $L(p,1)$, the parameter $p$ plays a
role analogous (but opposite) to the size $L$ of a torus in the
flat case.  The largest spherical 3-manifolds are $L (1,1)$ which
is the $3$-sphere and $L (2, 1)$ which is projective $3$-space.
Here we consider the generic case $p > 2$.  As $p$ increases, the
volume of the space $L (p, 1)$ decreases. For typical values of
$\nu$ the number of modes diminishes in rough proportion to $1 /
p$ according to the Weyl formula. However the $\nu = 2$ mode has
multiplicity $3$ for all $p > 2$. Thus the $\nu = 2$ mode has
proportionally higher weight in $L (p, 1)$ than in $\STR$, which
explains the excess of power on large angular scales (see
Fig.~\ref{fig1}). More precisely, after rescaling by the Weyl
formula's factor of $1 / p$, the $\nu = 2$ mode of $L (p, 1)$ has
an effective multiplicity of $3 / (1/p) = 3 p$, compared to the
$\nu = 2$ mode of $\STR$ which has multiplicity $9$.

The fact that the smallest nonzero eigenvalue of $L (p, 1)$ is
always $\nu = 2$ and has constant multiplicity $3$ for all $p > 2$
contrasts sharply to the behavior of the cubic $3$-torus of size
$L$, for which the smallest eigenvalue scales as $L^{-1}$.  This
contrasting behavior can be understood by realizing that as $p$
increases the space is becoming smaller in only one direction and
remains large in perpendicular directions: on large scales we see
a $2$-dimensional repartition of modes that are perpendicular to
the axis of the lens, and the relative weight of large scale as
compared to small scales is larger in lower dimensional spaces. An
informal examination of Fig.~\ref{fig1} suggests the very
conservative upper bound $p < 15$ when $\Omega_0 = 1.02$, for
consistency with WMAP's observation of a low quadrupole, keeping
in mind that this bound depends on $\Omega_0$. This constraint was
obtained on the basis of the angular power spectrum only, while a
comparison of the full correlation matrix is left for further
investigation. This bound on $p$ was set without any refined
statistical analysis mainly because the low $\ell$ part of the
spectrum becomes obviously inconsistent with WMAP data as $p$
increases. This conservative bound can be used to select the range
of a grid for further, more rigorous statistical investigations.
Let us also emphasize that our numerical results include all
components of the CMB, in particular the Doppler and integrated
Sachs-Wolfe contributions, the effects of which were detailed in
Ref.~\cite{rulw}.

Matching circles \cite{cornish98} occur when the distance $2 \pi /
p$ between two closest topological images is less than the
diameter of the last scattering surface, as summarized in
Table~\ref{table1}.

\begin{table}
\vskip0.25cm
\begin{tabular}{|c|c|c|c|}
\hline
$\Omega_0 - 1$ &
$\quad 0.01 \quad$ & $\quad 0.02 \quad$ & $\quad 0.03 \quad$ \\
\hline
$1/\sqrt{K} \UUNIT{(Gpc)}{}$ &
$48.4$ & $34.2$ & $27.9$ \\
$\CHN_\LSS \UUNIT{(rad)}{}$ &
$0.316$ & $0.442$ & $0.536$ \\
$\,\hbox{Vol} (< \CHN_\LSS) / \hbox{Vol}(\STR)$ &
$0.7 \%$ & $1.8 \%$ & $3.1 \%$ \\
\hline
detectability &
$p \geq 10$ & $p \geq 8$ & $p \geq 6$ \\
\hline
\end{tabular}
\caption{The first three rows give the curvature radius $1 /
\sqrt{K}$, the radius $\CHN_\LSS$ of the last scattering surface in
units of the curvature radius, and the volume of the observable
universe as a fraction of the whole universe.  The last row gives the
constraints on the order $p$ of the cyclic group so that the topology
scale is small enough to be detectable ($2 \pi / p < 2 \CHN_\LSS$).
All numbers are given for $\Omega_\Lambda = 0.7$, $\Omega_\MAT = 0.32
\pm 0.01$, and $h = 0.62$.}
\label{table1}
\end{table}

Figure~\ref{fig2} illustrates how the matching circles are diametrally
opposite but matched with a twist of $2 \pi / p$, simply because
Clifford translations twist and translate the same amount.  As $p$
increases, the map becomes more and more anisotropic with obvious
twisted structures (Fig.~\ref{fig3}).  The number of matched circles
is given by $N = p \CHN_\LSS /
\pi$ (see Fig.~\ref{fig1c}).

Recalling that double and linked action spaces combine the action of a
cyclic group $G' = Z_p$ with a more general group $G$, let us now
consider the effect of the $G$ factor.  As explained in
Ref.~\cite{wlu02}, the nearest repeating images typically depend only
on the $Z_p$ factor, that factor alone may generate matching circles,
and circle searching effectively reduces to the case of $L (p,
1)$. The $G$ factor, although typically irrelevant for circle
searching, may nevertheless affect the power spectrum.
Figure~\ref{fig1d} shows power spectra for $L (3, 1)$ ($G' = Z_3$, $G$
trivial), $L (12, 1)$ ($G' = Z_{12}$, $G$ trivial), $L (12, 5)$ ($G' =
Z_4$, $G = Z_3$), $L (60, 11)$ ($G' = Z_{12}$, $G = Z_5$), and $L
(132, 23)$ ($G' = Z_{12}$, $G = Z_{11}$).  Full understanding of the
power spectrum of a general $L (p, q)$ will require more numerical
investigation.

\section{Conclusion}

This article studied the imprint of a lens space topology on the
CMB. It provided simulated CMB maps and analyzed potential
detectability using the circle matching method.  The lens spaces $L
(p, 1)$ generically have a high quadrupole: when $\Omega_0 = 1.02$,
consistency with WMAP's quadrupole seems to imply $p < 15$, while for
the same value of $\Omega_0$ the LSS will intersect itself (and thus
produce potentially detectable matching circles) if and only if $p >
7$.


\end{document}